\documentstyle[12pt,aaspp4,flushrt,psfig]{article}
\begin{document}

\def\bq{\begin{equation}}
\def\eq{\end{equation}}
\def\bqy{\begin{eqnarray}}
\def\eqy{\end{eqnarray}}
\def\bqyn{\begin{eqnarray*}}
\def\eqyn{\end{eqnarray*}}
\def\bc{\begin{center}}
\def\ec{\end{center}}

%

\title{RXTE HARD X-RAY OBSERVATION OF A754:\\
CONSTRAINING THE HOTTEST TEMPERATURE COMPONENT AND THE
INTRACLUSTER MAGNETIC FIELD}

\author{
Azita Valinia\altaffilmark{1,2}, Mark J. Henriksen\altaffilmark{3},
Michael Loewenstein\altaffilmark{1,2}, Kurt Roettiger\altaffilmark{4},
Richard F. Mushotzky\altaffilmark{1}, and Greg Madejski\altaffilmark{1,2}
}


\altaffiltext{1}{Laboratory for High Energy
Astrophysics, Code 662, NASA/Goddard
Space Flight Center, Greenbelt, MD 20771; valinia@milkyway.gsfc.nasa.gov}
\altaffiltext{2}{Department of Astronomy, University of Maryland,
College Park, MD 20742}
\altaffiltext{3}{Physics Department, University of North Dakota, Grand Forks, 
ND 58202}
\altaffiltext{4}{Department of Physics and Astronomy, University of Missouri,
Columbia, MO 65211}

\begin{abstract}
Abell 754, a cluster undergoing merging, was observed in hard X-rays with the 
{\it Rossi X-ray Timing Explorer} (RXTE) in order to constrain its hottest temperature
component and search for evidence of nonthermal emission. Simultaneous modeling of RXTE
data and those taken with previous missions yields an average intracluster 
temperature of $\sim 9$~keV in the $1-50$~keV energy band. 
A multi-temperature component model 
derived from numerical simulations of the evolution of a cluster undergoing a merger 
produces similar quality of fit, 
indicating that
the emission measure from the very hot gas component is sufficiently
small that it 
renders the two models indistinguishable. No significant nonthermal emission
was detected. However, our observations set an upper limit of
$7.1 \times 10^{-14} \, {\rm ergs \, cm^{-2} \, s^{-1}}
\, {\rm keV}^{-1}$ (90\% confidence limit) 
to the nonthermal emission flux at 20~keV. Combining this result with the
radio synchrotron emission flux we find a lower limit of $0.2$~$\mu$G 
for the intracluster magnetic field. 
We discuss the implications of our results for the theories of magnetic
field amplifications in cluster mergers.  
\end{abstract} 
 
\keywords{galaxies: clusters: individual (Abell 754) ---  magnetic fields --- X-rays: clusters }

\newpage
%
 
\section{INTRODUCTION}

Clusters of galaxies are known to contain a large fraction of their mass in 
the form
of X-ray emitting gas in the temperature range $10^7-10^8$~K.  
The primary X-ray emission from the
intracluster medium (ICM) is thermal bremsstrahlung and
line emission due to this hot diffuse gas.
However, the possibility that some of the emission may be nonthermal
has been raised in the past (e.g. Bridle \& Feldman 1972; Harris \& 
Romanishin 1974). The most 
likely mechanism for nonthermal emission in the ICM is inverse
Compton (IC) scattering of cosmic microwave background (CMB) 
photons off relativistic
(GeV) intracluster electrons.
The presence of such energetic electrons
is even more likely within galaxy clusters that are known
to have significant radio emission, as it is suggested that the same
relativistic electron population produces the synchrotron emission.
As a result, the power law slope of the nonthermal X-ray
emission and that of the synchrotron radio emission are expected to
be the same.
Whereas very high energy electrons with short lifetimes are required for
synchrotron radio emission, lower energy electrons can scatter CMB 
photons up to X-ray energies (Felten and Morrison 1966).
For an electron with energy $E=\gamma m_e c^2$ where $\gamma (> 1000)$ is the
Lorentz factor, the final frequency of a CMB photon ($\nu_b$)
after scattering off the energetic electron will be (e.g. Pacholczyk 1970)
\bq
\nu_x={4 \gamma^2 \nu_b \over 3}.
\eq

Combining the X-ray inverse Compton and the radio synchrotron emission
fluxes, a lower limit to the magnetic field and an
upper limit to the energy density of relativistic electrons can be
derived {\it
independent} of equipartition or equal energy hypotheses.
Deriving a lower limit for the mean value of the magnetic field in the
ICM is particularly important for at least two reasons. First,
it affects total mass and
baryonic fraction estimates of the cluster because of the possibility
of extra magnetic pressure support. This has important consequences for the
underlying cosmological model and estimating the
value of the cosmological parameter $\Omega$.
For example, in the core of A2218 the X-ray mass
estimate, under the assumption that the gas is supported by thermal
pressure {\it alone}, is lower than the gravitational lensing
estimate by a factor of $2.5 \pm 0.5$.
It has been suggested that nonthermal pressure support is a plausible
explanation for the discrepancy (Loeb \& Mao 1994).
Second, detection of nonthermal emission and estimates of magnetic field
strength have important implications for cooling flows,
since the presence of the field can suppress conduction and
affect cooling of the gas.
Incidentally, we point out that it has been suggested that the 
luminous extreme-ultraviolet
(EUV) emission recently discovered in clusters of galaxies is also a result of
inverse Compton scattering of CMB radiation by
low energy cosmic ray electrons ($\gamma \sim 300$) in the intracluster
medium (Sarazin \& Lieu 1998).
These particles have been suggested to be a relic population of
cosmic rays produced by nonthermal activity over the lifetime of the cluster.
 
Aside from exploring nonthermal emission in clusters and deriving
limits on the ICM magnetic fields and relativistic
electron energy density, hard X-ray observations are needed
in the study of clusters that show evidence of extremely high
temperatures. There is ample evidence
that a component of extremely hot gas exists in some clusters (e.g. Abell 754)
that cannot be explained by the depth of the gravitational
potential (Henriksen \& Markevitch 1996; Roettiger, Stone \& Mushotzky 1998). 
Such hard X-ray emission could be the
result of hydrodynamical processes such as shocks.
Many clusters are believed to have formed through merger events as
predicted in hierarchical large-scale structure models. As a result,
due to the collision of subclumps and merging effects, very
high temperature shocked gas can be produced in the ICM.

There is evidence that Abell 754, a rich cluster of
galaxies at $z=0.054$, is a merger in progress.
The X-ray emission is offset significantly from the
bimodal galaxy distribution. Furthermore, the intracluster medium
is found to be nonisothermal (Henriksen \& Markevitch 1996). Its
X-ray morphology is asymmetric and not centered
on the galaxy distribution.
Our goal in observing this cluster with RXTE is two-fold. One is to
constrain the hottest temperature component of
its ICM using high energy data. This complements the
softer X-ray studies with
ASCA and provides further constraints for numerical
models of cluster evolution that, in turn, are necessary to
constrain the underlying physical processes in the ICM.
The second goal is to search for evidence of nonthermal emission in the
integrated spectrum from the cluster and to constrain the magnetic field and 
relativistic electron energy density in the ICM. 

The plan of this paper is as follows.
In \S 2, we review the derivation of magnetic field and relativistic electron
energy density limits from radio and X-ray observations. 
In \S 3, we describe the X-ray observations and the data reduction procedure. 
In \S 4, we present the results of the analysis. This includes results 
of spectral analysis using 
both isothermal and multi-temperature models (constructed from  numerical 
simulations), and constraints on nonthermal emission. 
In \S 5, we discuss the implications of these results and present
our conclusions. 

\section{Nonthermal X-ray Emission and Implications for the 
Intracluster Magnetic Field Strength}

Nonthermal X-ray emission in the intracluster medium is expected to arise 
as a result of the boosting of CMB photons to
X-ray energies from scattering off relativistic electrons.
This process is more likely in clusters with
diffuse radio halos,
since 
synchrotron radiation can be produced by
this same population of relativistic
electrons
traversing the ICM magnetic field.
From synchrotron emission alone, it is difficult to decouple the relativistic
electron density from the magnetic field strength without invoking the 
equipartition hypothesis.  
On the other hand, the electron spectrum can be derived from the X-ray emission
produced via inverse Compton scattering by energetic electrons of 
CMB photons, independent of the knowledge of the magnetic field. 
The radio synchrotron emission measurement can then be used to derive
limits on both the magnetic field strength and the electron energy density 
averaged along the line of sight. Detailed calculations are given, for example,
in Harris \& Romanishin (1974), Harris \& Grindlay (1979), 
and Rephaeli (1977a,b) with
many of the fundamental formulae given in Pacholczyk (1970). Here, 
following Henriksen (1998), we
review this technique from fundamental arguments.  

The number density of relativistic electrons in the ICM can be described
by a power law distribution
\bq
N(\gamma)=N_0\gamma^{-p}, \,\, \gamma_{\rm min} \le \gamma \le
\gamma_{\rm max}
\eq
where $\gamma$ is the electron Lorentz factor, and $N_0$ is the amplitude
coefficient. For this electron
distribution, the synchrotron and inverse Compton emission fluxes are
given by (using equations [6.36] and [7.31] of Rybicki and Lightman [1979])

\bq
F_s(\nu_r)={N_0 K_1 R_h^3 \over 3 D^2} B^{p+1 \over 2}
\nu_r^{1-p \over 2},
\eq
and
\bq
F_c(\epsilon_x)={N_0 K_2 R_h^3 \over 3 D^2} (kT)^{p+5 \over 2}
\epsilon_x^{1-p \over 2},
\eq
respectively, where $\nu_r$ is the frequency of the radio photons and
$\epsilon_x$ is the energy of the X-ray photons. $R_h$ is
the radius of the radio halo, $D$ is the distance to the cluster,
$B$ is the magnetic field
component perpendicular to the line of sight, $k$ is the Boltzmann constant,
and $T$ is the temperature of the CMB radiation.
$K_1$ and $K_2$ are given by

\bq
K_1={\sqrt{3} q^3 \over m c^2 (p+1)} \Gamma(p/4+19/12) \Gamma(p/4-1/12)
          (2 \pi m c/3 q)^{(1-p)/2},
\eq
and
\bq
K_2={8 \pi^2 r_0^2 \over h^3 c^2} J(p) \Gamma({p+5 \over 2})
         \zeta({p+5 \over 2}),
\eq
respectively, where $q$ is the electron charge, $m$ is the electron 
mass, $c$ is the speed of light, $r_0$ is the classical electron radius,  
$\Gamma$ is the Gamma function, $\zeta$ is the Riemann zeta
function defined by
\bq 
\zeta \equiv \sum_{n=1}^{\infty} n^{-s},
\eq  
and $J(p)$ is given by
\bq
J(p)=2^{p+3}{p^2+4p+11 \over (p+3)^2 (p+5) (p+1)}.
\eq  
These calculations assume that the electron distribution is isotropic and
that $\gamma_{\rm min}^2 << \nu_r / \nu_B << \gamma_{\rm max}^2$, where
$\nu_B \equiv qB/2\pi m_e c$ is the gyrofrequency in the magnetic field. 
The observed radio spectrum is given by a power law
function of the form
\bq
F_s(\nu_r)=A \nu_r^{-\alpha_r},
\eq
where $\alpha_r$ is the energy spectral index and is equal to the X-ray
spectral index $\alpha_x$. It is also related to the
electron power law index, $p$, via the relation $p=1+2 \alpha_r$. 
Since equation (4) can be independently solved for $N_0$, 
the relativistic electron energy density and magnetic field
component perpendicular to the line of sight can then be found from
\bq
\rho_e= N_0 \int_{\gamma_{\rm min}}^{\gamma_{\rm max}} (\gamma m c^2)
      \gamma^{-p} d\gamma,
\eq
and
\bq
B=\Big({3 A D^2 \over K_1 N_0 R_h^3}\Big)^{{2 \over p+1 }},
\eq
respectively. If the inverse Compton X-ray emission flux is only 
an upper limit, the
above equations yield an upper limit for $\rho_e$ and a lower limit for $B$.  

\section{OBSERVATIONS AND DATA ANALYSIS} 

\subsection{X-ray Observations}

Abell 754 was observed with the {\it proportional counter array} (PCA) 
and the {\it High Energy X-ray Timing Experiment} (HEXTE) instruments on board
{\it RXTE} during December of 1997 for a total duration of 70~ks. 
The PCA (Jahoda et al. 1996) has a total collecting area of 6500~${\rm
cm^2}$, an energy range of
$2-60$~keV, and energy resolution of $\sim 18\%$ at 6~keV. The collimator
field of view is approximately circular ($2^\circ$ diameter) with
FWHM of $1^\circ$. The HEXTE (Rothschild et al. 1998)
consists of two clusters, each
having a collecting area of 800~${\rm cm^2}$, an energy range of 
$15-250$~keV, energy resolution of 15\% at 60~keV, and a field of
view of $1^\circ$ FWHM. Furthermore, each cluster ``rocks'' along
mutually orthogonal directions to provide background measurements away
from the source.  

We used the most recent PCA background estimator program {\it pcabackest}
(version~2.0c; L7 model) provided by the 
{\it RXTE} GOF (Guest Observer Facility)
to estimate the background. In addition to the intrinsic instrument background,
and the cosmic X-ray background (CXB), additional counts are induced via 
SAA (Southern Atlantic Anomaly) passages.
A754 observations were performed during non-SAA orbits and  
therefore the activation induced background was estimated to be
zero by {\it pcabackest}.

We used the data taken when all 5 detectors of the PCA
were on and  the elevation angle from the limb of the Earth was
greater than  $10^\circ$. This reduced the good time interval for the purpose 
of analysis to $\sim 60$~ks. 
For this time duration, the background subtracted count rate over the
$3-20$~keV band was $41.04\pm0.0399\,{\rm counts\,s^{-1}}$. 
The background subtracted count rate over the $15-50$~keV energy band was
$0.3363\pm0.0728\,{\rm counts\,s^{-1}}$ and $0.3515\pm0.0607\,
{\rm counts\,s^{-1}}$ for HEXTE clusters A and B, respectively.
The total on-source integration time for the HEXTE detectors was 20~ks. 

X-ray images of A754 from the ROSAT PSPC archive indicate that
the radio
galaxy 26W20, a member of the cluster, lies in the field of view of RXTE. 
Since 
one of the goals of our investigation is to search for evidence of nonthermal 
emission in the ICM, we need to investigate the 
contribution of this galaxy to 
the integrated nonthermal emission from the cluster. 
The PSPC data, obtained in November 1992, 
reveal this source to be  
approximately $40^{'}$ 
from the center of A754, with a 
spectrum which is well-described as a power-law with a hydrogen column
density of 
$6.6 \pm 0.08 \times 10^{20}$ cm$^{-2}$, 
and photon index of $2.0 \pm 0.2$.  The $0.5-2$ keV flux  of this 
source is $2.5 \times 10^{-12}$ ergs cm$^{-2}$ s$^{-1}$, 
with a nominal 10\% error.  
With this spectrum, the $2-10$ keV flux would correspond to 
$\sim 3.8 \times 10^{-12}$ ergs cm$^{-2}$ s$^{-1}$, a factor of 
$\sim 30$ smaller than the integrated flux of A754 in the same band.  
Furthermore, at the position of the source in the field of view of PCA, the 
transmission is only about a third of the full response at the center of the
field of view. Therefore, we conclude that 
the contribution of 26W20 to the total integrated flux from A754 (particularly
in the $2-10$ band) is  negligible and does not affect the 
temperature determination of this cluster. On the other hand, the
$10-40$~keV flux of this galaxy would correspond to 
$\sim 3 \times 10^{-12} $ ergs cm$^{-2}$ s$^{-1}$. 
Since only about a third of this flux will get 
transmitted through the collimator, its contribution in the $10-40$~keV 
band will be $\sim 10^{-12}$ ergs cm$^{-2}$ s$^{-1}$.
We discuss its effect
on the detection of nonthermal IC radiation in \S 4.3.  

In addition to RXTE data, we have extracted {\it Ginga}~LAC
and {\it ASCA}~GIS2 spectra from archival observations of
this cluster with exposure times of
10 and 20~ks, respectively. The background subtracted count rate for the
LAC over 
the $3-18$~keV range was 
$20.24\pm0.0719\,{\rm counts\,s^{-1}}$, while for the  GIS2 over $1-10$~keV
was $1.687\pm0.0095\,{\rm counts\,s^{-1}}$. 
In \S 4, we present results from simultaneous fits 
to the spectra in the $1-50$~keV band obtained
from all 4 instruments.

\subsection{Radio Observations}

The 2.7 GHz data presented by Andernach et al. (1988) for the extended
radio source they associate with a possible diffuse ($12^{\prime}$) halo
in A754 has a flux density of 137~mJy. The synchrotron flux can then 
be written as 
$F_s=2.5 \times 10^{-12} \nu^{-1.3} \, {\rm ergs \, s^{-1}\, cm^{-2}
\, Hz^{-1}}$, where we have used the $\alpha_r=1.3$ reported
by Mills, Hunstead, \& Skellern (1978) and Jaffe  and Rudnick (1979) 
at 408 MHz and 610 MHz, respectively, since it is expected that the 
lowest frequency data should have the
least contamination from discrete radio sources. The 1.4 GHz data of 
Mills et al. (1978) and more recently 
Condon et al. (1998) NRAO/VLA Sky Survey (NVSS) show that at least
3 discrete sources can be resolved in the extend halo (these are
sources number 6, 7, and 8 in Figure~1a of Mills et al. [1978]). At least one
of these sources has been identified as a narrow angle tail radio galaxy
(NAT) by Zhao et al. (1989)
and Owen and Ledlow (1997). These 3 NVSS sources have flux densities of
$58.0$, $98.9$, and $20.4$~mJy, respectively, at 1.4~GHz. 
Using the Andernach et al. flux at 2.7 GHz and spectral index 
of $\alpha_r=1.3$ for the extended source, the expected flux density at
1.4~GHz is 322~mJy. Subtracting the flux density of the combined 
3 discrete sources from the total flux at 1.4~GHz yields a flux density
of approximately 145~mJy which may be attributed to the diffuse emission. 
The synchrotron flux for the diffuse radio emission can then be
written as 
\bq
F_s=1.1 \times 10^{-12} \nu^{-1.3} \, {\rm ergs \, s^{-1}\, cm^{-2}
\, Hz^{-1}}. 
\eq
We will use this flux relation in \S 4.3 to estimate a lower limit for the
magnetic field and an upper limit for the relativistic electron energy
density.

\section{RESULTS}  

\subsection{Isothermal Model}
We first attempt to fit the A754 spectrum
with the redshifted, optically thin collisional ionization
equilibrium plasma model of Raymond and Smith (Raymond \& Smith 1977). 
For comparison purposes, we first fit the spectra from RXTE PCA,
Ginga LAC, and ASCA GIS2 independently over the $3-10$~keV range 
(the overlapping energy band in all 3 instruments) with an 
isothermal model. 
Table~1 summarizes the fit parameters and the field of view of each
instrument. Note that A754 has a diameter of approximately
$0^\circ \! .5$, and fits within 
the field of view of all the above-mentioned
instruments. Reassuringly, the fitting results  
are consistent for all instruments within their respective error bars. 
We then fit the PCA/LAC/GIS2/HEXTE spectra simultaneously
in the $1-50$~keV energy band. 
Figure~1 shows the results of the fit, while 
Table~2 summarizes the best fit 
parameters. The fit is satisfactory and the derived
average temperature of $\sim 9$~keV is consistent
(within error bars) with the temperature derived from each individual
instrument. It is also consistent  
with the emissivity-weighted average temperature previously derived
using ASCA data (Henriksen and Markevitch 1996). 
The simultaneous PCA/HEXTE fit also yields similar results within error bars 
(although the size of the error bars are larger than the ones quoted 
in Table~2). 

\subsection{Multi-Temperature Model}

We have considered a multi-temperature model
based on 3D numerical simulations of A754
(Roettiger, Stone \& Mushotzky 1998).
The intracluster medium is evolved using an Eulerian
hydrodynamics code based on the Piecewise Parabolic Method (PPM)
and the collisionless dark matter is evolved via an N-body
particle mesh (PM) code.
In the simulations, two clusters are allowed
to merge under the influence of their mutual gravity, having been
given an initial relative velocity of 270 ${\rm km \, s^{-1}}$ parallel to the line
connecting the centers of the two subclusters and 100 ${\rm km \, s^{-1}}$ 
perpendicular to this direction. This results in a slightly
off-axis merger with an impact parameter of $\sim$120 kpc and final
impact velocity of $\sim 2500 \, {\rm km \, s^{-1}}$.
Roettiger et al. (1998) find that 
the epoch that most closely represents A754 occurs $\sim$0.3 Gyr
after the closest approach of the respective centers of mass.

Figure~2a shows the fractional distribution (by volume) of gas at a given
temperature within a
2 Mpc box centered on the X-ray emission. The cooler peak (at $\sim
6.5$~keV) represents
the initial primary cluster temperature.
All gas hotter than $\sim$7 keV has
been heated by the merger. In particular, the extended high temperature
tail is the result of a shock along the leading edge of one of the subclusters
to the NW.  Figure~2b shows the fractional distribution of the emission
measure of the gas at a
given temperature within the same region described for panel (a).
The two peaks in (a) have merged
into one near 7.5~keV. The effective emission-weighted temperature within
this region is $\sim$9 keV similar to that derived from observations.  

We implemented this model by summing a series of redshifted multi-temperature
Raymond-Smith plasma components of fixed temperature and 
relative emission measure following that
presented in Figure~2b, and fitted the data allowing
the column density $N_H$ and
metal abundance  -- assumed identical for all temperature components --
to vary. The goodness-of-fit 
($\chi^2=712.0$ for 636 d.o.f.) to the simultaneous RXTE/ASCA/Ginga data for
this model is slightly poorer but comparable to that of the isothermal
model. The best fit hydrogen column 
density and abundances for this model are also similar,
$0.10\pm0.02\times 10^{22}$ cm$^{-2}$ and $0.183\pm0.006$ solar, 
respectively. 
Essentially, this model is indistinguishable from the isothermal model
in moderate energy resolution integrated X-ray spectra.
This point can be further illustrated in Figure~3 where we have plotted the 
expected flux for the isothermal (dashed line) and multi-temperature model
(solid line), respectively, using the PCA response function. 
The two models are nearly indistinguishable up
to about 15~keV. Above 20~keV the difference in the hard tail widens. 
However,
at 40~keV, the multi-temperature model differs from the isothermal model
flux by approximately   
$ \sim 8 \times 10^{-14} \,{\rm ergs \, cm^{-2} \, s^{-1} \, keV^{-1}}$. 
This difference is smaller than 
$\sim 4 \times 10^{-13} \, {\rm ergs \, cm^{-2} \, s^{-1} \, keV^{-1}}$ 
or the fluctuations on Cosmic X-ray Background (CXB) per RXTE beam at 40~keV. 
Since any high temperature gas present in the cluster, as predicted in
numerical merger simulations,
has a small emission measure compared to that of the bulk of the gas, 
the integrated spectra of the two models are indistinguishable
with current instruments.

\subsection{Constraints on Nonthermal Emission and the Magnetic Field} 

To constrain the nonthermal emission in hard X-rays, we fit the RXTE data
(PCA and HEXTE) with an isothermal plus a power law model.
We assumed that the power law photon index in the X-ray model
is that obtained from the radio observations (\S 3.2).
To model the contribution of 26W20 in the field of view (40' away from
the center as discussed in \S 3.1), we included a fixed power law 
component of photon index 2
with a flux of $\sim 10^{-12} \,{\rm ergs \, cm^{-2} \, s^{-1}}$ in the
$10-40$ keV band. 
At the 90\% confidence limit, the combined thermal plus power law 
fit to the data gives only an upper limit of  
$\sim 4.9 \times 10^{-14} \,
{\rm ergs \, cm^{-2} \, s^{-1}}\, {\rm keV}^{-1}$ for the nonthermal 
flux at 20~keV.   
Integrated over the $10-40$~keV band, the flux upper limit is
$\sim 1.4 \times 10^{-12} \, {\rm ergs \, cm^{-2} \,s^{-1}}$.
At the 99.7\% confidence limit (or $3\sigma$ level), the above 
upper limits increase to 
$\sim 1.2 \times 10^{-13} $ and $ 3.5 \times 10^{-12} \,
{\rm ergs \, cm^{-2} \, s^{-1}}\, {\rm keV}^{-1}$, respectively.

The above result neglects the effect of CXB fluctuations in the spectrum.
However, these fluctuations have been determined to be about
8\% RMS of the mean CXB (Gruber 1998) per RXTE field of view. 
Using the energy flux (in units of ${\rm keV \,cm^{-2}\,s^{-1}\,keV^{-1}\,sr^{-1}}$)
\bq 
F(E)=7.9 E^{-0.29} \exp(-E/41.13\,{\rm keV}), \, 3\,{\rm keV}< E < 60\, {\rm keV}, 
\eq 
for the CXB (Gruber 1992), 
the fluctuation amplitude amounts to $6.5 \times 10^{-14} \,
{\rm ergs \, cm^{-2} \, s^{-1}}\, {\rm keV}^{-1}$ per RXTE field of view 
at 20 keV. 
Since this value is comparable to the nonthermal upper limit 
derived for A754, the effect of CXB fluctuations on modeling of the data 
cannot be neglected. 
To determine a more accurate upper limit, we modeled the fluctuation spectrum
by equation (13) and fixed its amplitude in our spectral modeling. 
At the 90\% confidence limit, a positive full amplitude fluctuation 
(i.e. 8\% of the CXB) yields an upper limit 
of $\sim 3.6 \times 10^{-14} \,
{\rm ergs \, cm^{-2} \, s^{-1}}\, {\rm keV}^{-1}$ for the nonthermal 
flux at 20~keV and
an integrated flux of 
$\sim 10^{-12} \, {\rm ergs \, cm^{-2} \,s^{-1}}$
in the 10-40~keV band. At the 99.7\% confidence limit, the above upper 
limits increase to
$\sim   10^{-13}$ and $2.8 \times 10^{-12} \, {\rm ergs \, cm^{-2} \,s^{-1}}$,
respectively. 
At the 90\% confidence limit, a negative amplitude fluctuation 
yields $\sim 7.1 \times 10^{-14} \,
{\rm ergs \, cm^{-2} \, s^{-1}}\, {\rm keV}^{-1}$ and $\sim 2 \times 
10^{-12} \, {\rm ergs \, cm^{-2} \,s^{-1}}$ for the flux at 20~keV and 
the $10-40$~keV integrated flux, respectively. 
At the 99.7\% confidence limit, the above upper limits increase to
$\sim 1.6 \times  10^{-13}$ and $4.4 \times 10^{-12} \, 
{\rm ergs \, cm^{-2} \,s^{-1}}$, respectively.

To find a lower limit on $B$ and an upper limit on $\rho_e$, we solve 
equation (4) for $N_0$ by using the upper limit inverse Compton flux over
the $10-40$~keV energy band. We use the 90\% confidence limits to derive
limits on $B$ and $\rho_e$.  
Since our upper limit to the $10-40$~keV flux
is in the range $(1-2) \times 10^{-12} \, {\rm ergs \, cm^{-2} \,s^{-1}}$, 
an upper limit of 
$(1.4-2.8) \times 10^{-3}\, {\rm cm^{-3}}$ is derived for $N_0$.  
The detection of nonthermal X-ray photons at
energy $E_x$ primarily samples electrons
of energy $\gamma= (3 E_X/4 h \nu_b )^{1/2}$
where $\nu_b$   
is the CMB photon frequency, and $m_e$ and $c$ are the electron rest 
mass and the speed of light,
respectively. 
Integrating equation (10) over the $10-40$~keV range using the upper limit
flux derived over this band, 
we find an upper limit of $(1.1-2.2) \times 10^{-15} \,{\rm ergs \, cm^{-3}}$
for the relativistic electron energy density in the intracluster medium.
Equation~(11) yields a lower limit of $0.2-0.27\,\mu$G for the magnetic field
strength. 
Note that the electron energy range
over which we integrated equation (10) corresponds to radio emission in
the range $16-64$~MHz ($E=\gamma m_e c^2= \sqrt{\nu_r/c_\circ B}$;
$c_\circ=6.27 \times 10^{18})$, which is the frequency of maximum synchrotron
emission (Pacholczyk 1970). Our calculations, therefore,
assume that the radio spectrum
extends down to 16~MHz with the same spectral index. 
We note that from simultaneous analysis of HEAO-1 and ASCA
data, Henriksen (1998) finds a lower limit of $0.29$~$\mu$G for this cluster.

\section{DISCUSSION and CONCLUSIONS}

We have
presented X-ray observations and data analysis of Abell 754, a cluster in
the process of a merger event. In addition to the newly obtained RXTE PCA/HEXTE data, 
we obtained archival ASCA GIS and Ginga LAC
data of this cluster and simultaneously fitted 
the spectra from all 4 instruments. We found that the integrated spectrum of
the whole cluster in the $1-50$~keV energy band can be fitted well with an isothermal
collisional ionization equilibrium plasma model (Raymond \& Smith 1977) of 
temperature $\sim 9$~keV, consistent within the error bars 
with previous
temperature measurements of this cluster. 

We then fitted the data with a multi-temperature 
model derived from numerical simulations of two 
merging subclusters with conditions
similar to those found in Abell 754 (Roettiger et al. 1998). The average    
emission weighted temperature from the simulations was $\sim 9$~keV, similar
to that found from observations. In this model, the emission measure from the 
very hot gas, shock-heated in the merger, is relatively small. 
As a result, this model produces a fit to the data of similar quality
to that of the isothermal model. Calculation of the resultant 
spectrum from the two models indicates that the two spectra are nearly indistinguishable
below $\sim 15$~keV and it is only above 20~keV that the hard X-ray tails from
the two models depart. However, the difference between the two models at 40~keV
is less than $\sim 10^{-13} \,{\rm ergs \, cm^{-2} \, s^{-1} \, keV^{-1}}$, 
lower than the
amplitude of the CXB fluctuations at that energy per RXTE field of view.
Hence, the two models are indistinguishable with the obtained RXTE spectrum.  

To search for evidence of nonthermal emission from the cluster, we added a power law
to the isothermal model with the photon index fixed at the value obtained for the
radio emission power law. No significant nonthermal emission was found. However, 
the RXTE data implies an upper limit (90\% confidence limit) of  
$\sim (3.6-7.1) \times 10^{-14} \, {\rm ergs \, cm^{-2} \, s^{-1}}\, 
{\rm keV}^{-1}$ for the nonthermal flux at 20~keV, 
where the lower and upper numbers are those derived assuming
a full positive and negative CXB fluctuation amplitude as discussed in \S 4.3, 
respectively.  
Using the upper limit flux integrated over the $10-40$~keV band and the
radio emission data from Andernach et al. (1988), we find an upper limit of
$(1.1-2.2) \times 10^{-15} \,{\rm ergs \, cm^{-3}}$
for the relativistic electron energy density and a lower limit of 
$0.2-0.27\,\mu$G for the magnetic field strength in the intracluster medium, 
where the given ranges represent the inclusion of positive and negative
CXB fluctuations. 

The magnetic field would have to be two orders of magnitude greater than
this lower limit in order for the average magnetic pressure to be comparable
to the average thermal pressure in the hot gas. In regions of higher than
average density where the field would be expected to be compressed to
higher values, the thermal pressure is enhanced by a comparable factor.
Therefore, there is no evidence for dynamically important
magnetic fields in A754.

The upper limit to the magnetic field derived
assuming energy equipartition of the magnetic
field with the relativistic electrons is $0.17-0.23\,\mu$G, below  
(or comparable) to the lower limit field strength derived from combining 
radio and X-ray observations (depending on the unknown level of 
the CXB fluctuations
for that region of the sky). If it is the case that the
equipartition field is lower than that observed, one interpretation
is that since Abell 754 is undergoing a merger,
magnetic fields are being amplified in the merger process and the 
ensuing turbulence. This interpretation has been confirmed in numerical
simulations of cluster mergers where the field is observed to be amplified
by several factors in localized regions (Roettiger, Stone, \& Burns 1998). 
Unfortunately, it is difficult to draw any firm conclusions regarding
the relation between the equipartition value and the actual strength of
the magnetic field in A754 based on the available data. 
Higher resolution VLA 90 cm
radio images of the extended emission and higher spatial resolution X-ray
images (e.g. with AXAF) are needed to make more definitive statements.  

There have been several lines of 
speculation on magnetic field generation in the 
ICM. One school of thought
is that galactic wakes power a dynamo (e.g. Jaffe 1980; Roland 1981).
However, as shown by Goldman \& Rephaeli (1991) and De Young (1992), galactic
wakes do not adequately produce the observed fields. It has been suggested 
that more powerful sources such as cluster mergers could be responsible (e.g. 
De Young 1992). Since both magnetic
field amplification and reacceleration of energetic particles is expected 
in cluster mergers from the resulting
shocks and turbulence, large scale radio halos
are likely to be formed around these clusters. However, while the magnetic
reconnection time in the intracluster medium,
\bq
t_{rec} \approx 2 \times 10^9 ({\epsilon \over 0.1})^{-1} ({l_t 
\over 1\,{\rm kpc}}) ({B \over\mu{\rm G}})^{-1}({n_p \over
0.1 \, {\rm cm^{-3}}})^{1/2} \, {\rm yr},
\eq
is long (where the reconnection proceeds with an average velocity of $\epsilon v_A$
and $\epsilon \approx 0.1$; Soker \& Sarazin 1990), 
the inverse Compton energy loss time given by 
\bq
t_{IC}={\gamma m_e c^2 \over {4 \over 3} \sigma_T c \gamma^2 U_{CMB}}
\approx 4.8 \times 10^8 ({\gamma \over 4800})^{-1}\, {\rm yr} 
\eq
for an electron of energy $\sim 20$~keV ($\gamma \approx 4800$) 
is comparatively short. This may be the underlying reason for radio halos
to be a transient phenomenon and the fact that they are usually associated
with cluster mergers or dynamically young clusters (Tribble 1993).  
This association has also been noted by Edge et al. (1992) 
and Watt et al. (1992). 

The ultimate origin of the magnetic fields in clusters is still debatable.
Suggestions have been made that magnetic field and relativistic particles
in clusters originate within cluster radio sources and
are dispersed into the intracluster
medium during merger events (Tribble 1993, Harris et al. 1993). Finally, we note
that the recent discovery of EUV emission in clusters of galaxies and its
possible origin as inverse Compton scattering of CMB photons off low 
energy cosmic ray electrons indicate that old radio halos should be present
in a large number of clusters, albeit at very low frequencies that are not
detectable from Earth. 
 
\acknowledgements

This research has taken advantage of HEASARC and LEDAS archival data bases.

%

{}
\clearpage

%

\begin{center}
Figure Captions
\end{center}

\figcaption[]
{(a) Simultaneous fit to PCA/HEXTE/GIS/LAC data over the $1-50$~keV band
with an isothermal model (best fit parameters given in Table~2). 
(b) Same as (a) except the unfolded spectrum is shown. 
\label{fig1}}

\figcaption[]
{(a) Fractional distribution (by volume) of gas at a given temperature within a
2 Mpc box centered on the X-ray emission. The cooler peak represents
the initial primary cluster temperature. All gas hotter than $\sim$7 keV has
been heated by the merger.
(b) Fractional distribution (weighted
by differential emission measure n$^2$dV) of gas at a given temperature
within the same region described in (a). The two peaks in (a) have merged
into one near 7.5 keV. The effective emission-weighted average
temperature within this
region is $\sim$9 keV.
\label{fig2}} 

\figcaption[]
{Expected spectrum of isothermal (dashed line) and multi-temperature (solid
line) models described in \S 4.1 and \S 4.2 using the PCA response matrix.}

\clearpage 

\begin{figure}
\vspace{5in}
\includegraphics{f1a_a754.ps}
\end{figure}

\clearpage

\begin{figure}
\vspace{5in}
\includegraphics{f1b_a754.ps}
\end{figure}

\clearpage

\begin{figure}
\vspace{5in}
\includegraphics{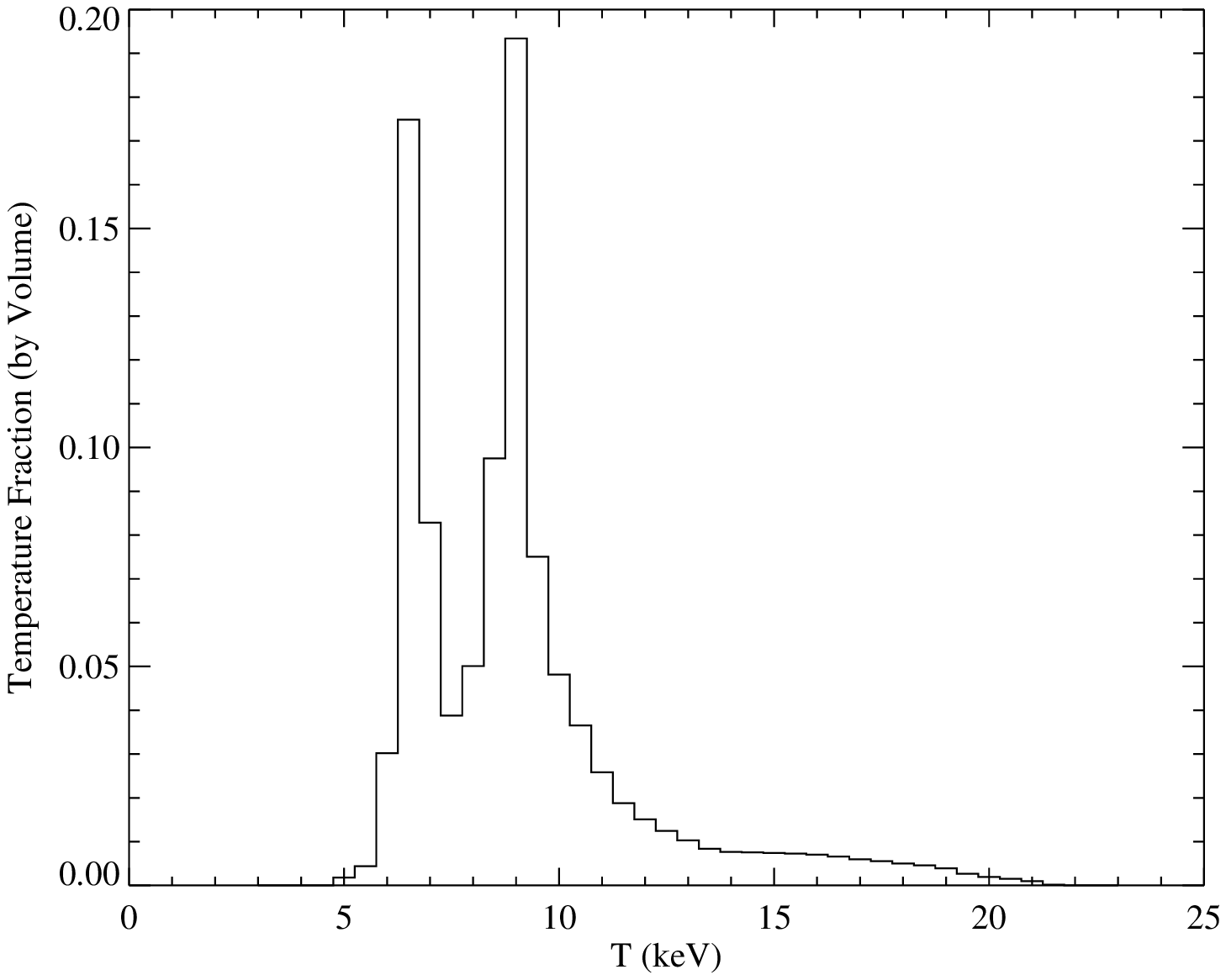}
\end{figure}

\clearpage

\begin{figure}
\vspace{5in}
\includegraphics{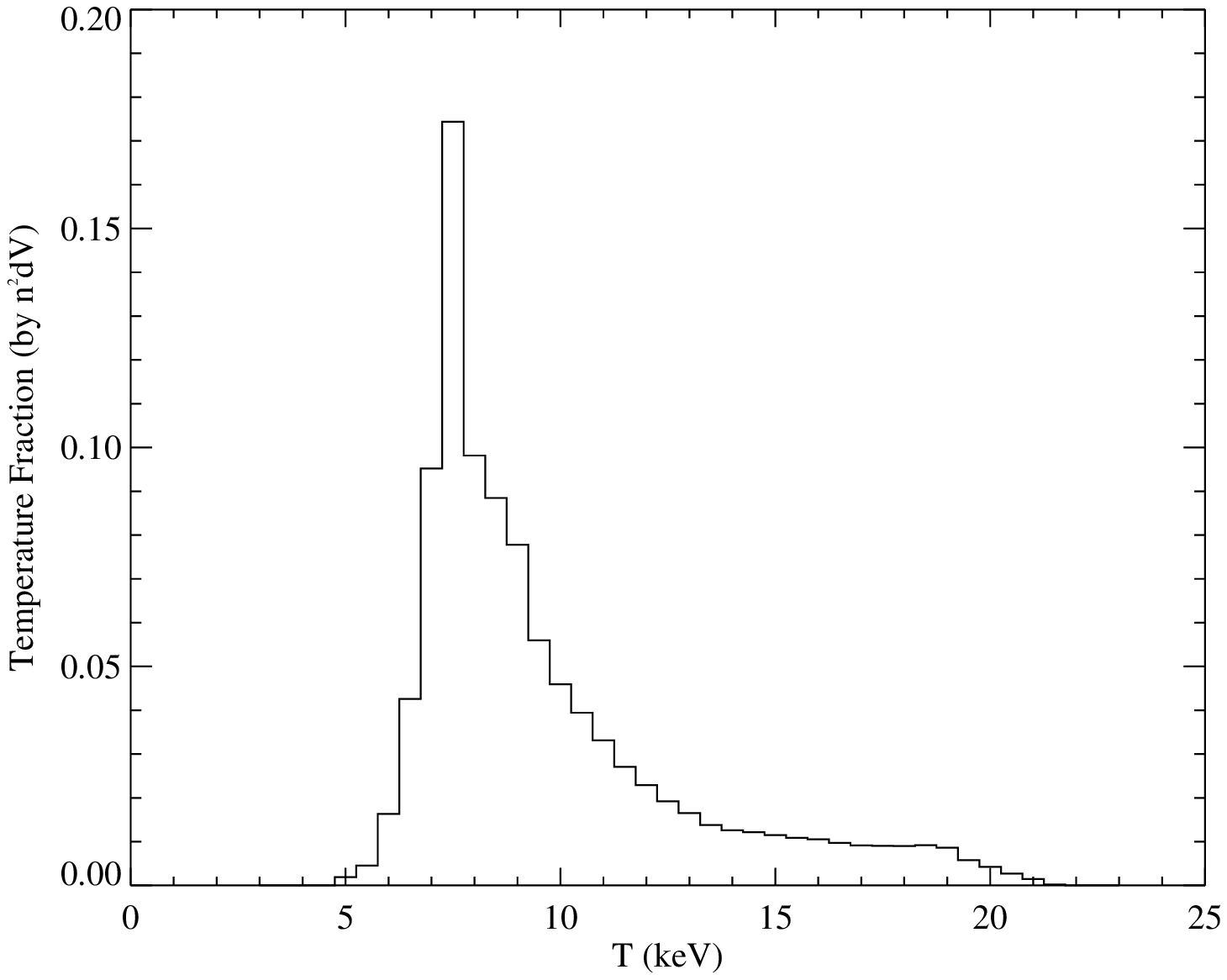}
\end{figure}

\clearpage

\begin{figure}
\vspace{5in}
\includegraphics{f3_a754.ps}
\end{figure}

\clearpage

\begin{deluxetable}{ccccc}
\footnotesize
\tablecaption{Isothermal Model Fit Over $3-10$~keV Energy Range}
\tablewidth{500pt}
\tablehead{
\colhead {Instrument} &
\colhead {FOV (FWHM) } & 
\colhead {kT} &
\colhead {Abundance} &
\colhead {$\chi^2(\nu)$ }}
\startdata

RXTE PCA\tablenotemark{a}  &    $1^\circ$  & $9.06\pm0.13$  & 
$0.177\pm0.012$  &  21.8(15) \nl
Ginga LAC  & $1^\circ \times 2^\circ$  &  $9.26\pm0.28$  & $0.18\pm 0.03$
&  11.2(9)  \nl  
ASCA GIS  &  $40'$ diameter    & $10.93\pm2.1$ &  $0.23\pm0.09$ &  327.8(333) \nl

\enddata
\tablenotetext{a} {Due to the residual Xenon L-edge at $\approx 4$~keV in the 
detector's response, 0.004 ${\rm counts \, s^{-1}}$ systematic error was added
to the spectrum to obtain a reduced $\chi^2$ below 2 for the fit.}  
\end{deluxetable}

\clearpage

\begin{deluxetable}{lc}
\footnotesize
\tablecaption{Best Fit Parameters for 
the Isothermal Model over $1-50$~keV Energy Range:
Simultaneous Fits for ASCA/GINGA/PCA/HEXTE}
\tablewidth{500pt}
\tablehead{
\colhead {Parameter} &
\colhead {Value }} 
\startdata

$N_H\tablenotemark{a}$ ${\rm atom \, cm^{-2}}(10^{22})$  &  $0.08\pm0.02$ \nl
$kT$  & $8.98\pm0.06$   \nl
Abundance  &  $0.174\pm0.006$  \nl
RXTE PCA Normalization &  $0.0997\pm 0.0004$ \nl
RXTE HEXTE A/PCA cross normalization\tablenotemark{b} & $0.72\pm0.1$  \nl
RXTE HEXTE B/PCA cross normalization\tablenotemark{b} & $0.61\pm0.12$  \nl
ASCA GIS/PCA cross normalization\tablenotemark{b} &  $1.04\pm0.02$  \nl
Ginga LAC/PCA cross normalization\tablenotemark{b} & $0.84\pm0.009$  \nl
$\chi^2$(d.o.f)  & 659.9(634) \nl

\enddata
\tablenotetext{a} {Galactic value in the direction of A754 is 0.045.}
\tablenotetext{b} {Instrument normalization can be obtained by 
multiplying this number by the PCA normalization.} 
\end{deluxetable}


\clearpage

\begin{thebibliography}{}

\bibitem[ander]{ander}
Andernach, H., Tie, H., Sievers, A., Reuter, H.-P., Junkes, N., \&
Wielebinski, R. 1988, A\&AS, 73, 265 

\bibitem[brid]{brid} 
Bridle, A. H. \& Feldman, P. A. 1972, Nature Phys. Sci., 235, 168  

\bibitem[condon]{condon} 
Condon, J. J., Cott, W. D., Greisen, E. W., Yin, Q.F., 
Perley, R. A., Taylor, G. B., \& Broderick, J.J. 1998 AJ, 115, 1693

\bibitem[edge]{edge}
Edge, A. C., Stewart, G. C., \& Fabian, A. C., 1992, MNRAS, 258, 177 


\bibitem[felten]{felten}
Felten, F. E., \& Morrison, P., 1966, ApJ, 146, 686 

\bibitem[gruber92]{gruber92}
Gruber, D. E., 1992, in {\it The X-ray Background}, eds. Barcons, X. \& Fabian, A. C.,
pp. 44-53  

\bibitem[gruber]{gruber}
Gruber, D. E., 1998, private communication 

\bibitem[Harr]{harr}
Harris, D. E., \& Romanishin, W. 1974, ApJ, 188, 209  

\bibitem[harris]{harris}
Harris, D. E., Stern, C. P., Willis, A. G., Dewdney, P. E., 1993, AJ, 105,
769

\bibitem[henrika]{henrika}
Henriksen, M. J, 1998, PASJ, in press

\bibitem[henrik]{henrik}
Henriksen, M. J, \& Markevitch, M. L. 1996, ApJL, 466, 79 

\bibitem[jaffe]{jaffe}
Jaffe, W., and Rudnick, L., 1979, ApJ, 233, 453

\bibitem[Jahoda et al]{Jahoda}
Jahoda, K., Swank, J. H., Giles, A. B., Stark, M. J., Strohmayer,
T.,
Zhang, W., \& Morgan, E. H. 1996, Proc. SPIE, 2808, 59

\bibitem[loeb]{loeb}
Loeb, A., \& Mao, S., 1994, ApJL, 435, 109  

\bibitem[Mills]{mills}
Mills, B., Hunstead, R., Skellern, D., 1978, MNRAS, 185, 51

\bibitem[owen]{owen}
Owen, F. N., \& Ledlow, M. J., 1997, ApJS, 108, 41

\bibitem[pach]{pach}
Pacholczyk, A. G., 1970, Radio Astrophysics (San Francisco: Freeman) 

\bibitem[ray]{ray}
Raymond, J. C., \& Smith, B. W., 1977, ApJS, 35, 419  

\bibitem[repha]{repha}
Rephaeli, Y., 1977a, ApJ, 212, 608

\bibitem[rephb]{rephb}
Rephaeli, Y., 1977b, ApJ, 218, 323

\bibitem[roett1]{roett1}
Roettiger, K., Stone, J. M., \& Burns, J. O. 1998, ApJ, submitted

\bibitem[roett]{roett}
Roettiger, K., Stone, J. M., \& Mushotzky, R. F., 1998, ApJ, 493, 62  

\bibitem[roth]{roth}
Rothschild, R. E., Blanco, P. R., Gruber, D. E., Heindl, W. A.,
MacDonald, D. R., Marsden, D. C., Pelling, M. R., Waune, L., R.,
\& Hink, P. L. 1998, ApJ, 496, 538 

\bibitem[ryb]{ryb}
Rybicki, G., \& Lightman, A., 1979, Radiative Processes in Astrophysics, John
Wiley and Sons, Inc. 

\bibitem[sar]{sar}
Sarazin, C. L., \& Lieu, R. 1998, ApJL, 494, 177 

\bibitem[sok]{sok}
Soker, N., \& Sarazin, C., 1990, 348, 73 

\bibitem[trib]{trib}
Tribble, P. C., 1993, MNRAS, 263, 31  

\bibitem[Watt]{watt}
Watt, M. P., Ponman, T. J., Bertram, D., Eyles, C. J., Skinner, G. K., \&
Willmore, A. P., 1992, MNRAS, 258, 738 
\end{thebibliography}
\end{document}